\documentclass[preprintnumbers,amsmath,amssymb]{revtex4}
\usepackage{graphicx}
\usepackage{dcolumn}
\usepackage{bm}

\usepackage[pdftex]{color}

\usepackage{epsfig}

\begin{document}

\newcommand{\Eqref}[1]{(\ref{#1})}
\newcommand{\tf}[2]{\textstyle\frac{#1}{#2}}
\newcommand{\df}[2]{\displaystyle\frac{#1}{#2}}

\title{Interactions and Focusing of Nonlinear Water Waves}

\author{Harihar Khanal, Stefan C. Mancas}
\email{harihar.khanal@erau.edu, stefan.mancas@erau.edu}

\affiliation{Department of Mathematics, Embry-Riddle Aeronautical University,\\ Daytona Beach, FL. 32114-3900, U.S.A.}

\author{Shahrdad Sajjadi}
\email{sajja8b5@erau.edu}
\affiliation{Center for Geophysics and Planetary Physics, Embry-Riddle Aeronautical University}


\begin{abstract}

The coupled cubic nonlinear Schr\"odinger (CNLS) equations are used to
study modulational instabilities of a pair of nonlinearly interacting
two-dimensional waves in deep water. It has been shown that the full
dynamics of these interacting waves gives rise to localized
large-amplitude wavepackets (wave focusing). 
In this short letter we attempt to verify this result numerically using a Fourier spectral method 
for the CNLS equations. 

\end{abstract}

\maketitle

\section{Introduction}
Extremely large size waves (commonly known as freak, rogue or giant 
waves) are very common in the open sea or ocean and they pose major hazard to mariners. 
As early as 1976, Peregrine \cite{Peregrine} suggested that in the region of oceans where there is a strong current present, freak waves can form when action is conncentrated by reflection into a caustic region. 
A variable current acts analogously to filamentation instability in laser-plasma interactions \cite{NicholasSajjadi1,NicholasSajjadi2}.
Freak waves are very steep and is a nonlinear phenomena, hence they cannot
be represented and described by a linear water wave theory. Zakharov \cite{Zakharov09} in 2009 
has noted that in the last stage of their evolution, their steepness becomes
`infinite', thereby forming a `wall of water'.
However, before  such an instant in time, the steepness is higher 
than one for the limiting Stokes wave and before 
breaking the wave crest reaches three to four (sometimes even more) times higher 
than the crests of neighboring waves. The freak wave is preceded by a deep trough 
appearing as a `hole in the sea'. On the other hand, a characteristic life time 
of a freak wave is short, typically ten of wave periods or so. For example, if the 
wave period is fifteen seconds, then their life time is just few minutes. 
Freak wave appears almost instantly from a relatively calm sea. It is, therefore,
easy to appreciate that such peculiar features of freak waves cannot be 
explained by means of a linear theory. Even the focusing of ocean waves is a 
preconditions for formation of such waves.


It is now quite common to associate appearance of freak waves with the 
modulation instability of Stokes waves. This instability (known as 
the Benjamin--Feir instability) was first discovered by 
Lighthill 
\cite{Lighthill} 
and the detail of theory was developed independently by Benjamin and Feir 
\cite{BenjaminFeir} 
 and by Zakharov 
 \cite{Zakharov68}. 
Zakharov showed slowly modulated weakly nonlinear Stokes wave can be described by 
nonlinear Shr\"odinger equation (NLSE) and that 
this equation is integrable \cite{ZakharovShabat} and is just the 
first term in the hierarchy of envelope equations describing packets of 
surface gravity waves. The second term in this hierarchy was calculated 
by Dysthe 
\cite{Dysthe}.

Since the pioneer work of Smith 
\cite{Smith}, many researchers attempted (both theoretically
or numerically) to explain the freak wave formation by NLSE. Among diverse results obtained by them 
there is one important common observation which has been made by all, and that is, 
nonlinear development of modulational instability leads to 
concentration of wave energy in a small spatial region. This marks the  
possibility for formation of freak wave. 
Modulation instability leads to decomposition of initially 
homogeneous Stokes wave into a system of envelope solitons, or more strictly 
quasi-solitons \cite{ZakharovKuznetsov, Zakharovetal}. This state can be called `solitonic turbulence', 
or `quasisolitonic turbulence'. 
 
In this letter, we consider the problem of a single solition in a homogeneous media, being subjected 
to modulational instability which eventually leads to formation of a system of soliton. We 
will show that the supercritical instability leads to maximum formation of soliton, concentrated
in a small region. In going through subcritical instability the solitons coagulate
to early stages of supercritical instability. Moreover, we investigate the full
dynamics of nonlinearly interacting deep water waves subjected to
modulational/filamentation instabilities, and we find that random
perturbations can grow to form inherently nonlinear water wave
structures, the so called freak waves, through the nonlinear
interaction between two coupled water waves.  The latter should be
of interest for explaining recent observations in water wave
dynamics.

\section{Numerical Approach} 

In a pioneering work, a theory for
the modulational instability of a pair of two-dimensional
nonlinearly coupled water waves in deep water, as well as the
formation and dynamics of localized freak wave packets was presented \cite{shuk,r3}. The two wave packets were investigated in the context of nonlinear optics by \cite{Agr,Ber}, in Bise-Einsten condensates by \cite{Kas,Kour}, in transmission by \cite{Bil}, and in plasmas by many other authors \cite{In, Mc,Shukla,Kourakis,Vlad}.

Following \cite{shuk}, we consider the two-dimensional CNLS equations in the following form
\begin{eqnarray}\label{CNLSE1}
i \biggl( \frac{\partial A}{\partial t} + C_x \frac{\partial
A}{\partial x} + C_y \frac{\partial A}{\partial y} \biggr) +
\alpha \frac{\partial^2 A}{\partial x^2} + \beta \frac{\partial^2
A}{\partial y^2} + \gamma \frac{\partial^2 A}{\partial x \partial
y}- \xi \,|A^2|  A - 2 \zeta \,|B|^2  A = 0
\, ,
\end{eqnarray}
and
\begin{eqnarray}\label{CNLSE2}
i \biggl( \frac{\partial B}{\partial t} + C_x \frac{\partial
B}{\partial x} - C_y \frac{\partial B}{\partial y} \biggr) +
\alpha \frac{\partial^2 B}{\partial x^2} + \beta \frac{\partial^2
B}{\partial y^2} - \gamma \frac{\partial^2 B}{\partial x \partial
y}- \xi \,|B^2|  B - 2 \zeta \,|A|^2 B  =
0, 
\end{eqnarray}
where $A$ and $B$ are the amplitudes of the slowly varying wave
envelopes. 
The $x$ and $y$ components of the group velocity are
given respectively by
$$C_x=\omega k/2\kappa^2\quad {\rm and}\quad C_y=\omega\ell/2\kappa^2$$ 
and the group velocity dispersion
coefficients are 
$$\alpha=\omega(2 \ell^2-k^2)/8\kappa^4,\quad
\beta=\omega (2k^2-\ell^2)/8\kappa^4\quad{\rm and}\quad\gamma=-3\omega \ell k/4\kappa^4.$$ 
Also, the nonlinearity coefficients (as in \cite{r3}) are given by
$\xi=\omega \kappa^2/2$ and
$$\zeta=\omega(k^5-k^3\ell^2-3k\ell^4-2k^4\kappa+2k^2
\ell^2\kappa+2\ell^4\kappa)/2\kappa^2(k-2\kappa).$$ 

Here $k$ and $\ell$ are wavenumbers and $\omega$ is the wave frequency. They are related by 
$\omega =\sqrt{g\kappa}$ (the dispersion relation for deep water waves \cite{Karpman}) with $g$ the acceleration due to gravity 
and $\kappa$ the wavenumber norm given by $\kappa \equiv \sqrt{k^2 + \ell^2}$.  
For detail description of the formulation of the problem, we refer to the original works \cite{r3,shuk}.

The nonlinear strongly coupled system of equations (\ref{CNLSE1}) and (\ref{CNLSE2}) will be computed using a fast numerical algorithm based on the spectral method \cite{Beylkin, Trefethen} which is explained below. 
\subsection{Fourier Spectral Method}
We first noticed that by  letting $S=A+B$ and $D=A-B$, the system  (\ref{CNLSE1}) and (\ref{CNLSE2}) becomes symmetric,
obtained from (\ref{CNLSE1}) and (\ref{CNLSE2})
\begin{eqnarray}
i\left(\frac{\partial S}{\partial t} + C_x \frac{\partial S}{\partial x} + C_y \frac{\partial D}{\partial y}\right) + \alpha \frac{\partial^2 S}{\partial x^2}
+ \beta \frac{\partial^2 S}{\partial y^2}
+ \gamma \frac{\partial^2 D}{\partial x\partial y}
=g(S,D) 
\label{eq1}\\
 i\left(\frac{\partial D}{\partial t} + C_x \frac{\partial D}{\partial x} + C_y \frac{\partial S}{\partial y}\right) + \alpha \frac{\partial^2 D}{\partial x^2}
+ \beta \frac{\partial^2 D}{\partial y^2}
+ \gamma \frac{\partial^2 S}{\partial x\partial y}
=G(D,S) 
\label{eq2}
\end{eqnarray}
where 
\begin{equation}\label{eq5a}
G(u,v)= \frac18\left[(\xi+2\eta)\left(|u+v|^2+|u-v|^2\right)u+(\xi-2\eta)\left(|u+v|^2-|u-v|^2\right)v \right]
\end{equation}

Then, we reduce the above system of PDEs (\ref{eq1})--(\ref{eq2}) into a system of ODEs using the Fourier transform of $u(x,y)$ which is defined by
\begin{equation}\label{eq4}
\mathcal{F}(u)(k_x,k_y)=\widehat{u}(k_x,k_y)=\frac{1}{2\pi}\int_{-\infty}^{\infty}\int_{-\infty}^{\infty}e^{-i(k_xx+k_yy)}u(x,y)\, dx\,dy,
\end {equation}
with the corresponding inverse
\begin{equation}\label{eq5}
\mathcal{F}^{-1}(\widehat{u})(x,y)=u(x,y)=\frac{1}{2\pi}\int_{-\infty}^{\infty}\int_{-\infty}^{\infty}e^{i(k_xx+k_yy)}\widehat{u}(k_x,k_y)\, dk_x\,dk_y.
\end {equation}
The function $\widehat{u}(k_x,k_y)$ can be interpreted as the amplitude density of $u$ for wavenumbers $k_x$, $k_y$. 
Now, we take the Fourier transform of both (\ref{eq1}) and (\ref{eq2}) 
 as
\begin{equation}\label{eq6}
i\frac{d\widehat{S_t}}{dt}-\left(k_xC_x+\alpha k_x^2 + \beta k_y^2\right)\widehat{S} -k_y\left(C_y+\gamma k_x\right)\widehat{D}=\widehat{G(S,D)},
\end{equation}
\begin{equation}\label{eq7}
i\frac{d\widehat{D_t}}{dt}-\left(k_xC_x+\alpha k_x^2 + \beta k_y^2\right)\widehat{D} -k_y\left(C_y+\gamma k_x\right)\widehat{S}=\widehat{G(D,S)},
\end{equation}

Letting  $k_xC_x+\alpha k_x^2 + \beta k_y^2=p$ and $k_y\left(C_y+\gamma k_x\right)=r$  (\ref{eq6}) and (\ref{eq7}) can be written in the matrix form as 
\begin{equation}\label{eq8}
i\frac{d}{dt}\left(
\begin{array}{l}
\widehat{S}\\
\widehat{D}
\end{array}
\right)
=\left(
\begin{array}{cc}
p & r\\
r & p
\end{array}
\right)\left(
\begin{array}{l}
\widehat{S}\\
\widehat{D}
\end{array}
\right)
+ 
\left(
\begin{array}{l}
\widehat{G(S,D)}\\
\widehat{G(D,S}
\end{array}
\right)
\end{equation}

Computing the eigenvalues 
and eigenvectors 
the solution to (\ref{eq8})  
can be written as 
\begin{eqnarray}\label{eq9}
\left(
\begin{array}{l}
\widehat{S}\\
\widehat{D}
\end{array}
\right)
&=&\frac12\left(
\begin{array}{cc}
e^{-i\lambda_1 t} +e^{-i\lambda_2 t}  & -e^{-i\lambda_1 t} +e^{-i\lambda_2 t} \\
-e^{-i\lambda_1 t} +e^{-i\lambda_2 t}  & e^{-i\lambda_1 t} +e^{-i\lambda_2 t} 

\end{array}
\right)\left(
\begin{array}{c}
\widehat{S}(0)\\
\widehat{D}(0)
\end{array}
\right)\nonumber\\
&& +\, \frac12\int_0^t \left(
\begin{array}{c}
e^{i\lambda_1 \tau} \left(\widehat{G(S,D)}-\widehat{G(D,S)} \right)  \\
e^{i\lambda_2 \tau} \left( \widehat{G(S,D)}+\widehat{G(D,S)}\right)
\end{array}
\right) d\tau
\end{eqnarray}
with $\lambda_1=k_xC_x+\alpha k_x^2 + \beta k_y^2-k_y\left(C_y+\gamma k_x\right)$ and $\lambda_2=k_xC_x+\alpha k_x^2 + \beta k_y^2+k_y\left(C_y+\gamma k_x\right)$.

We exploit the symmetry of the nonlinear function $G$ from (\ref{eq5}) in developing a numerical procedure to solve the system of ODEs (\ref{eq8}). 

\subsection{Spatial discretization (Discrete Fourier Transform)}

We discretize the spatial domain $\Omega=[-L/2,L/2] \times [-L/2,L/2]$ into $n \times n$ uniformly spaced grid points $X_{ij}=(x_i,y_j)$ with $\Delta x=\Delta y= L/n$, $n$ even, and $L$ the length of the rectangular mesh $\Omega$. Given $u(X_{ij})=U_{ij},\, i,j=1,2,\cdots,n$, we define the 2D Discrete Fourier transform (2DFT) of $u$ as
\begin{equation}\label{eq10}
\widehat{u}_{k_xk_y} = \Delta x \Delta y \sum_{i=1}^{n} \sum_{j=1}^{n} e^{-i(k_xx_i+k_yy_j)}U_{ij},\,\,\,\,\,\, k_x,k_y=-\frac{n}{2}+1,\cdots, \frac{n}{2}
\end{equation}
and its inverse 2DFT as
\begin{equation}\label{eq11}
U_{ij} = \frac{1}{(2\pi)^2} \sum_{k_x=-n/2+1}^{n/2} \sum_{k_y=-n/2+1}^{n/2} e^{i(k_xx_i+k_yy_j)}\widehat{u}_{k_xk_y},\,\,\,\,\,\, i,j=1,2,\cdots, n.
\end{equation}
In equation (\ref{eq10}) and (\ref{eq11}) the wavenumbers $k_x$ and $k_y$, and the spatial indexes $i$ and $j$,
take only integer values.\\


\subsection{Temporal discretization}
We solve the initial value problem of the ODE system (\ref{eq8}) using 
the classical fourth order Runge-Kutta (RK4) method and exact treatment for the linear part  \cite{Beylkin}.

Given $t_\textrm{max}$, we discretize the time domain $[0,t_\mathrm{max}]$ with equal time steps of size $\Delta t$ with  $t_n=n\Delta t , \, n=0,1,2,\cdots$, and define $S^n=S(x,y;t_n)$ and $D^n=D(x,y;t_n)$.
Initializing $\widehat{S}^n=\widehat{S}(t_n)$ and $\widehat{D}^n=\widehat{D}(t_n)$, we compute the Fourier transforms of the nonlinear terms 
$
\mathcal{F}\left(G\left(\mathcal{F}^{-1}(\widehat{S}^n), \mathcal{F}^{-1}(\widehat{D}^n) \right)\right)$ and $\mathcal{F}\left(G\left(\mathcal{F}^{-1}(\widehat{D}^n), \mathcal{F}^{-1}(\widehat{S}^n) \right)\right)$,  
and advanced the ODE (\ref{eq8}) in time
with time step $\Delta t$ using the explicit RK4  for the nonlinear part, together with an exact solution for the linear part as shown in (\ref{eq9}).

\subsection{Simulation setup}  
The numerical code for the above procedure is implemented in FORTRAN 90 and executed on a Linux cluster. 

The initial profiles for $A$ and $B$ were taken as Gaussians, 
\begin{eqnarray}
A(x,y;0)=(A_0+random(O(10^{-3}/\kappa))e^{-\sigma(x^2+y^2)}\\
B(x,y;0)=(B_0+random(O(10^{-3}/\kappa))e^{-\sigma(x^2+y^2)}\
\end{eqnarray}

In the simulations reported here,  we used the parameter values 
$\theta_0=\pi/6$, $g=9.81$, $w=0.56$, $k=0.33$, $A_0=0.1/\kappa$, $B_0=A_0,0$, $\sigma=1,0$, $L=2$ and a grid of $256\times 256$ nodes in the computational domain $[-1,1]\times[-1,1]$ with  the time step size $\Delta t =0.01$. 

For each simulation we monitor the energies $Q_A(t)$ and $Q_B(t)$,  calculated as 
\begin{equation}
Q_A(t)=\int_{-\infty}^{\infty}\int_{-\infty}^{\infty} |A(x,y;t)|^2 \, dx\,dy=\sum_{i=1}^{n}\sum_{j=1}^{n}  |A_{ij}|^2  \Delta x \Delta y
\end{equation}
\begin{equation}
Q_B(t)=\int_{-\infty}^{\infty}\int_{-\infty}^{\infty} |B(x,y;t)|^2 \, dx\,dy=\sum_{i=1}^{n}\sum_{j=1}^{n}  |B_{ij}|^2  \Delta x \Delta y
\end{equation}  

Observing a finite energy will reveal stability of a solution. As soon as the solution becomes unstable, the energy diverges. When the solution dissipates the energy approaches to zero. 

\begin{figure}[ht]
\begin{center}
{\includegraphics[width=2.40in,height=1.68in]{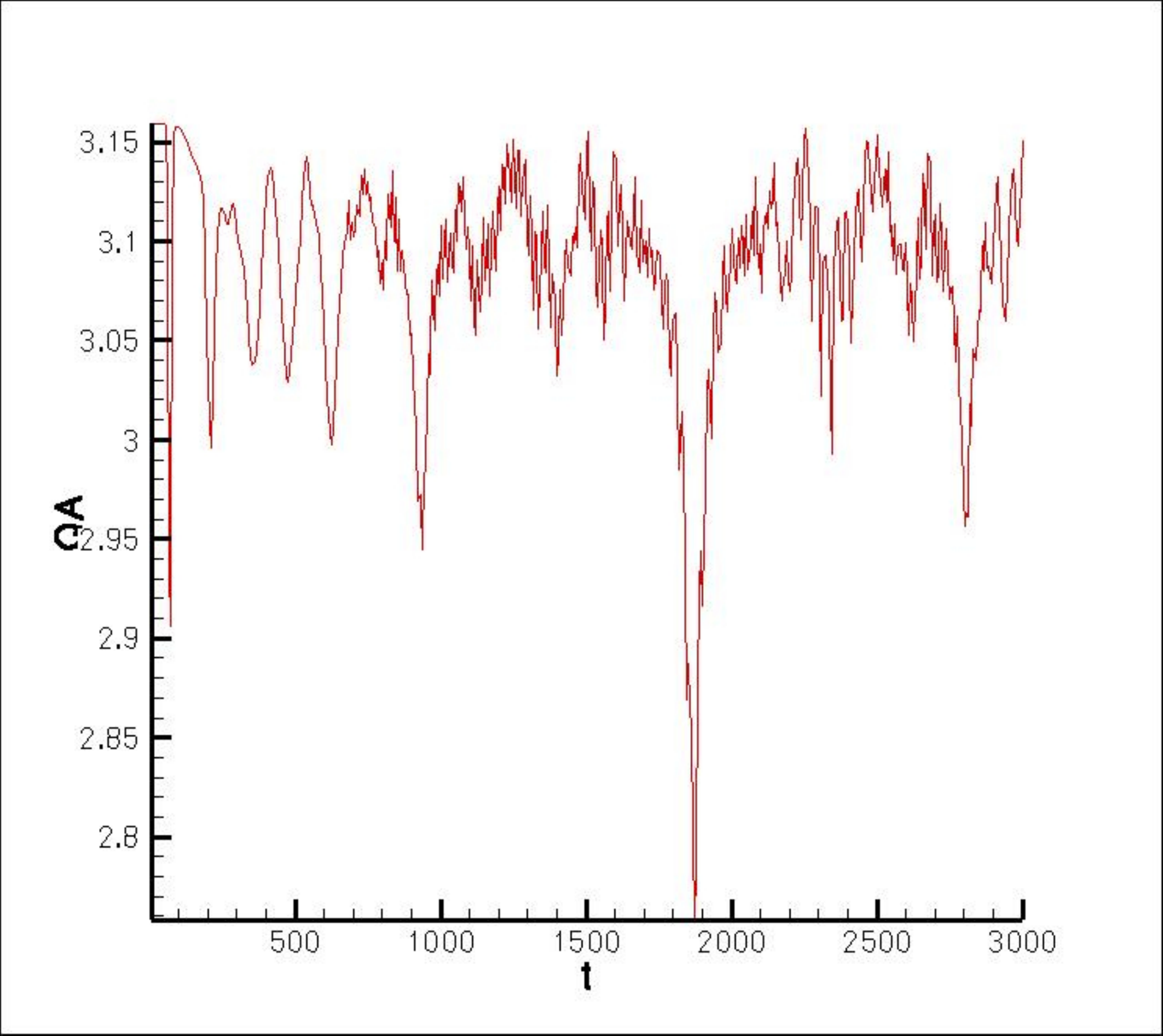}}
\hspace*{0.5in}
{\includegraphics[width=2.40in,height=1.68in]{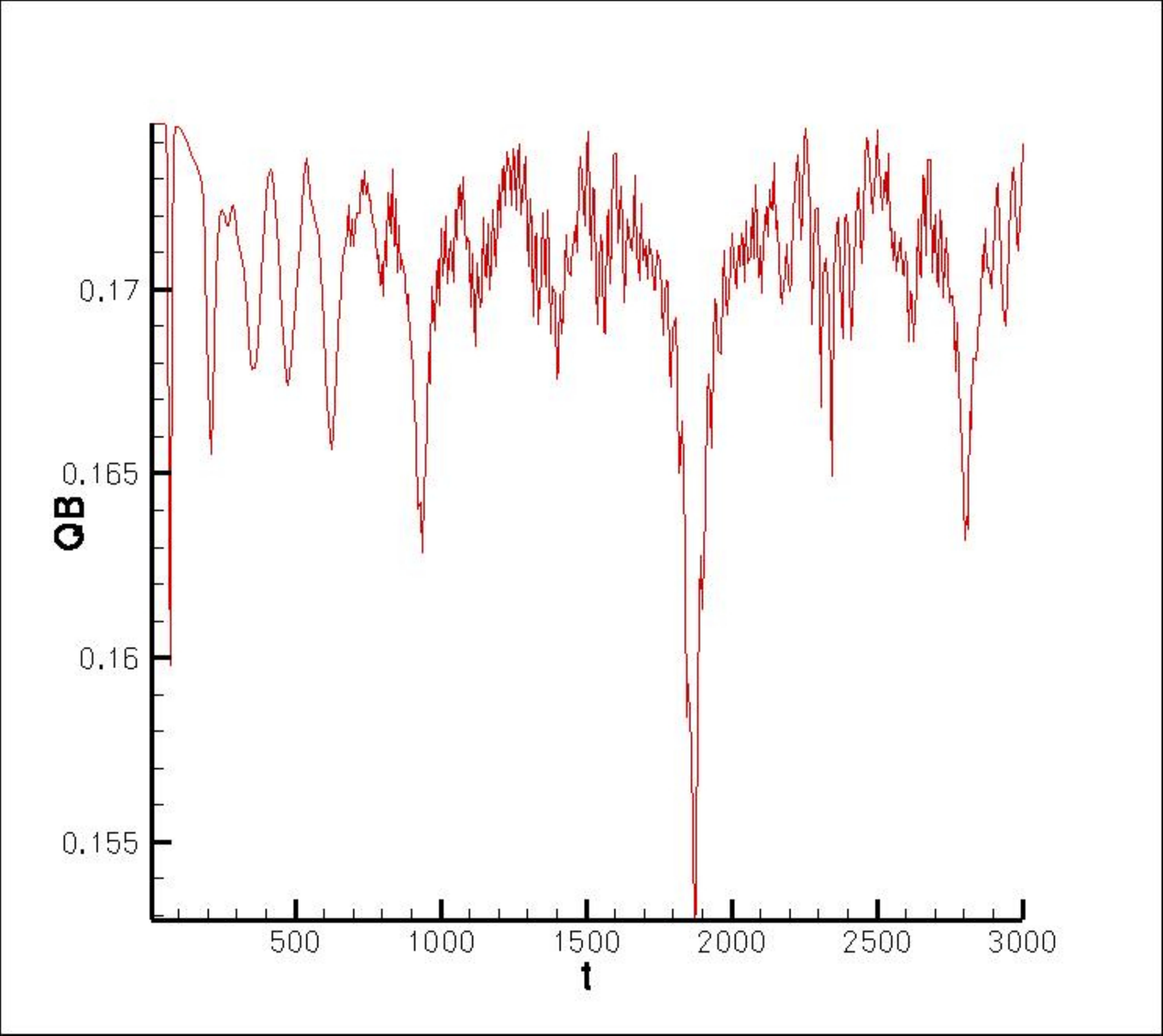}}
\label{fig1}
\end{center}
\caption{Energy evolution $Q_A(t)$ (left) and $Q_B(t)$ (right) in a typical 3000 min. simulation.}
\end{figure}

\section{Results}

\begin{figure}[ht]
\begin{center}
{\includegraphics[width=2.40in,height=1.68in]{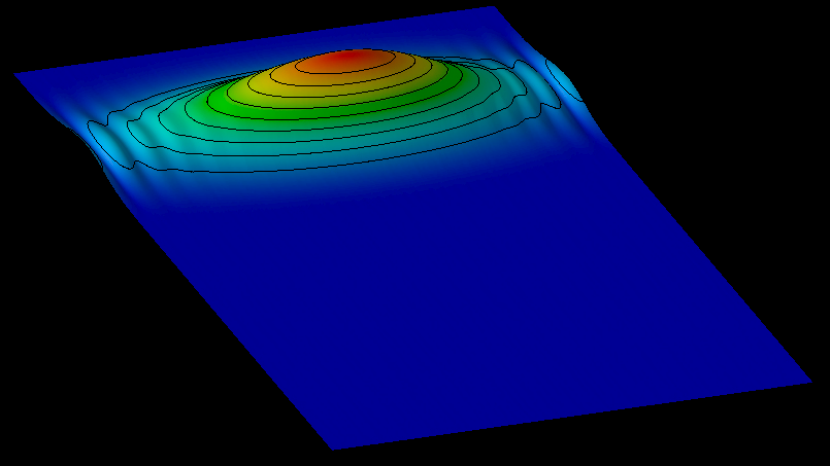}}
\hspace*{0.5in}
{\includegraphics[width=2.40in,height=1.68in]{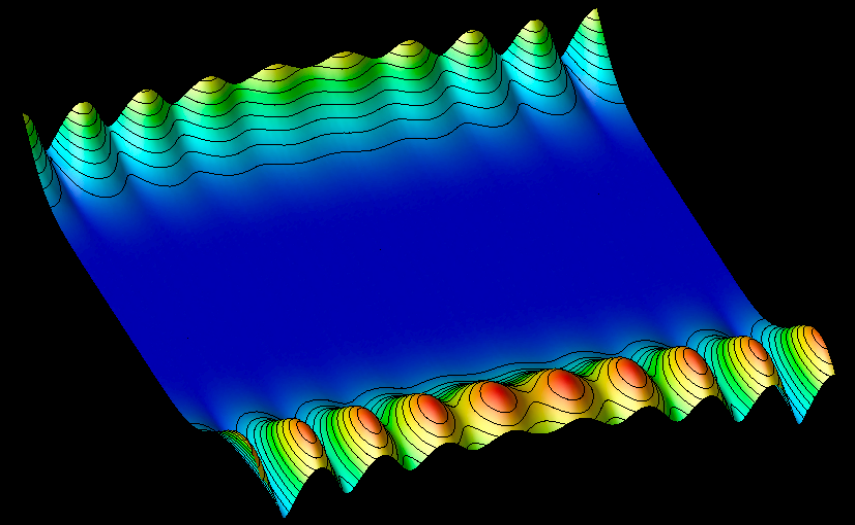}}
{\includegraphics[width=2.40in,height=1.68in]{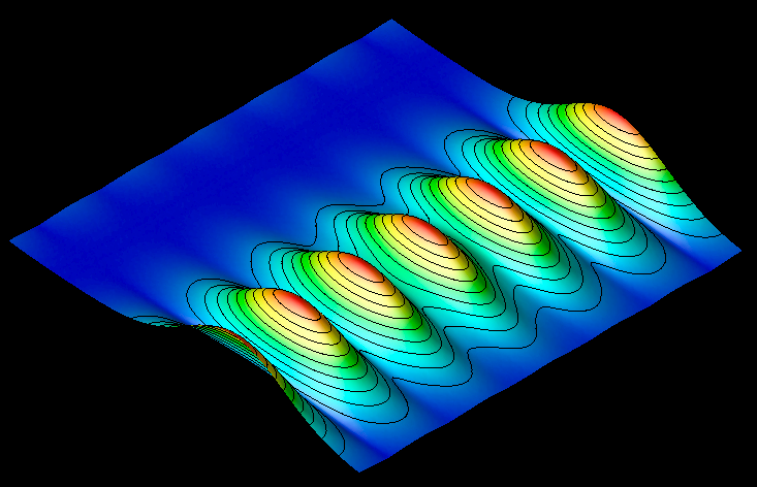}}
\hspace*{0.5in}
{\includegraphics[width=2.40in,height=1.68in]{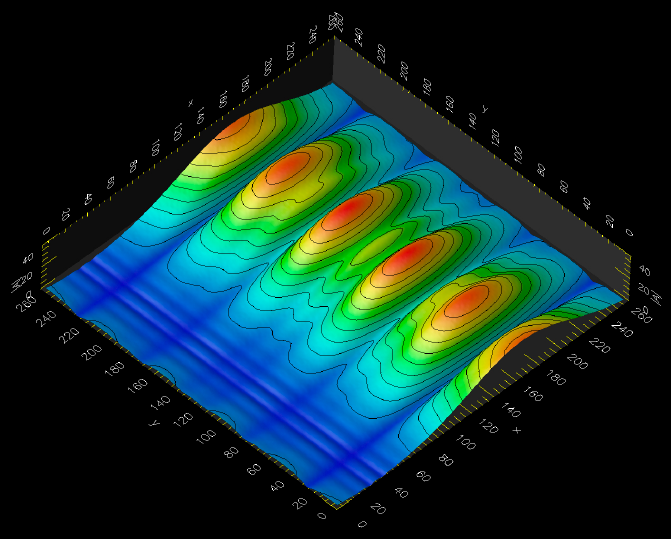}}
\hspace*{0.05cm}
{\includegraphics[width=2.40in,height=1.68in]{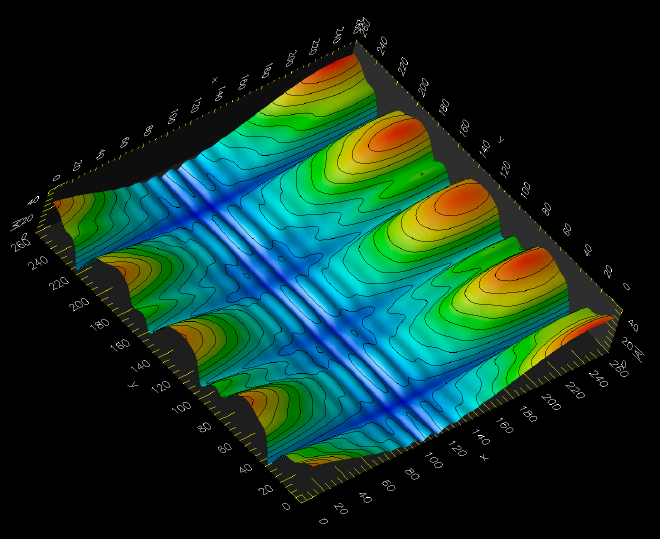}}
\hspace*{0.5in}
{\includegraphics[width=2.40in,height=1.68in]{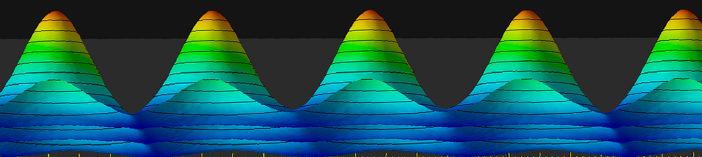}}
\label{fig2}
\end{center}
\caption{ The interaction between two waves, with equal initial
amplitudes $|A|=|B|=0.1\,\kappa^{-1}$ which are propagating at an
angle of
$\theta=\pi/6$.  A low-amplitude noise equal to
$10^{-3}/\kappa$ is added to the initial amplitude in order to
enhance the modulation instability.}
\end{figure}

The results presented in this paper
represent a preliminary study on dynamics of interacting nonlinear water
waves. The problem considered here comprises the dynamics of nonlinear interacting water
wave packets through solving the coupled system of equations (\ref{CNLSE1}) and (\ref{CNLSE2}).

The results that are shown in Fig. 2 are all in dimensional units, where
the two interacting waves initially have the amplitude
$A=B=0.1/\kappa+{\rm ran}$, with {\rm ran}  representing a random low-amplitude 
noise, equal to $10^{-3}/\kappa$, in order  to enhance instability. The
results shown represent different time steps (starting on the 
left-hand panel and going downwards) for $t=300/\omega $, $t=600/\omega $, $t=900/\omega $
then (right-hand panel) $t=1200/\omega $, $t=1500/\omega $; the last figure on the 
right-hand panel is at the same time as that above it but plotted 
from a different prospective reflecting the maximum growth rate in the $y$ direction.
For our simulations we have taken typical data from ocean waves \cite{Hasselmann76}. 
The waves $A$ and
$B$ in Fig. 2, then have the initial amplitudes
$|A|=|B|=0.1/\kappa\approx 3$ meters. From these figures, we see at
$t=1500/\omega$ ($\approx 2680 $ seconds) that wave $A$ focuses
as a localized wave packets with a maximum amplitude of $\approx
0.35/\kappa\approx 10$ meters. We remark for considerable period 
after the initial step, waves $A$ and $B$ are qualitatively the same
(with $|A|>|B|$) before the nonlinear wave-wave interactions set in 
which results to wave break-up. 


In summary, 
we presented a numerical procedure to solve CNLS equations describing 
modulational instabilities of a pair of nonlinearly interacting
two-dimensional waves in deep water. 
The simulation results of the full dynamical system reveals that even
waves that are separately modulationally stable can, when
nonlinear interactions are taken into account, give rise to novel
behavior such as the formation of large-amplitude coherent wave
packets with amplitudes several times the
initial waves. This behavior is quite different from that of a
single wave (the case for the original Benjamin-Feir
instability) which disintegrates into a wide spectrum of waves. These
results are relevant to the nonlinear instability arising from
colliding water waves thereby producing large-amplitude oceanic freak waves.

\end{document}